\documentstyle[aps]{revtex}

\begin{document}

\draft

\title{Structure and Stability of an Amorphous Metal}
\author
{Oscar Rodr\'{\i}guez de la Fuente$^{\ast}$ and Jos\'{e} M. Soler}
\address{
Departamento de F\'{\i}sica de la Materia Condensada
and Instituto Nicol\'{a}s Cabrera, C-III \\
Universidad Aut\'{o}noma de Madrid, 28049 Madrid, Spain.}
\maketitle

\begin{abstract}
Using molecular dynamics simulations, 
with a realistic many-body embedded-atom potential,
and a novel method to characterize local order, 
we study the structure of pure nickel 
during the rapid quench of the liquid and in the resulting glass. 
In contrast with previous simulations with pair potentials,
we find more crystalline order and fewer icosahedra for slower quenching
rates, resulting in a glass less stable against crystallization.
It is shown that there is not a specific amorphous structure, only
the arrest of the transition from liquid to crystal,
resulting in small crystalline clusters immersed in an amorphous
matrix with the same structure of the liquid.

\end{abstract}
The detailed knowledge of the atomic structure is essential to
understand the special properties of amorphous materials. 
There seems to be general agreement in that the basic process 
underlying the glass transition is the arrest of structural
relaxations \cite{ang}, but it is not clear to what extent this leads to
new structural units.
Very different structural models have been proposed \cite{gas},
and some of them assume 
that the glass structure is essentially
different from those of the liquid and the crystal.
For close-packed systems, the icosahedron 
has been largely proposed as a characteristic 
configuration, which becomes predominant during liquid quenching.
Another example is the polycluster model \cite{bak}, 
in which the amorphization proceeds by the nucleation and growth 
of amorphous clusters within the liquid.

However, like in liquids and defective solids, the lack of
atomic periodicity makes the experimental determination 
of the structure a largely unsolved problem.
Therefore, molecular dynamics simulations 
offer an invaluable complementary tool,
although the simulated times are generally much shorter
than the experimental quenches.
A common strategy is to study model systems with
simple interactions, hoping
that they will describe qualitatively the
properties of real glasses.
Thus, systems simulated with pair
potentials show an increase of icosahedra
during liquid cooling \cite{yon,kon,jun1,ste}.
However, the structural properties depend strongly on the 
type of interatomic potential used \cite{par}, and
some simulations indicate that the formation
of icosahedra decreases when many-body effects or nonadditivity 
are included \cite{pos}. 
Therefore, it is questionable to what extent these results
can be extrapolated to real glasses.

An alternate strategy is to focus on systems,
like pure metals, whose natural relaxation times are so short 
that they can be actually simulated in the limit between glass 
formation and crystallization. 
Within this approach we show in this work that,
in the case of a pure metal, there is
no specific amorphous structure
different from that of the liquid and the crystal, but
only the arrest of the transition from one to the other.
We also study the stability of the glass and, in contrast
with recent works \cite{yon,jun2}, show that it is more stable
when quenched faster.

Although the experimental quenching rates required to 
form their glasses are extremely fast, the possible 
applications of pure amorphous metals make their study
a field of much current interest \cite{roj,jud}.
In this work we have simulated the formation of pure amorphous
nickel using a realistic embedded-atom potential \cite{daw}.
It belongs to a broad class of potentials \cite{daw,nor} 
with the general functional form:
\begin{equation}
U = \sum_{i<j} V(r_{ij}) + \sum_i F\left( \sum_{j\neq i} \rho(r_{ij}) \right)
\end{equation}
The first term typically represents the repulsion of 
overlapping ionic cores. The second term is the energy
of immersing an atom in the electron density produced
by its neighbors, and represents the metallic bond.
This type of potential has
proven to reproduce very well the basic structural and dynamical
properties of solids, surfaces, defects,
liquids \cite{sad} and glasses \cite{bro}
of transition and noble metals.
Thus, although our calculations were performed for the specific
case of nickel, we expect that the qualitative conclusions
will remain valid for a wide set of pure metals.
To make the simulation more realistic, we use constant 
temperature and pressure molecular dynamics, thus allowing
energy and volume fluctuations which may be critical to
the resulting dynamics close to the glass transition.
In particular, it is well known that the glass formation
tendency of many materials depends strongly on the pressure \cite{ang2}.

The simulated system consists of 6912 atoms in a cubic cell
with periodic boundary conditions, initially equilibrated
at 2000 K, well above the melting temperature.
The Newton equations were integrated using the Verlet
algorithm with a time step of 5 fs. 
The temperature and pressure were fixed
by rescaling the atomic velocities and positions at every time step.
The pressure was kept always zero, while the temperature was 
reduced linearly from 2000 K to 100 K at different cooling rates,
from $8 \times 10^{12}$ to $10^{14}$ K/s.
The systems were finally equilibrated at 100 K during 160 ps
to prove their stability.
Following Stillinger \cite{sti},
to analize the structure at arbitrary temperatures
without the vibrational noise,
we always relax first the system to its closest local minimum
by a conjugate gradient energy minimization.
To characterize the atomic environments, we have 
generalized the method of Steinhardt {\it et al.} \cite{ste},
based on the rotationally invariant moments:
\begin{equation}
Q_l=\frac{1}{N_n} \left(\sum_{ij}^{N_n} P_l(\cos \theta_{ij}) \right) ^{1/2}
\end{equation}
where $N_n$ is the number of nearest neighbors of the atom
(closer than the first minimum in the radial distribution function),
$\theta_{ij}$ are the angles formed by two neighbors (with the 
reference atom in the vertex), and $P_l$ are Legendre polynomials.
The $Q_l$'s, with $0 \le l \le 10$, may be considered as the 
components of a vector.
If the euclidean distance from this vector to that of the fcc, hcp
or icosahedral environments is less than certain tolerances,
the atom has that environment distorted in some degree \cite{tol}.
Varying these tolerances in a wide range did not change 
qualitatively any of the results reported.
For the embedded atom potential used, the fcc and hcp structures are 
essentially degenerate and they are present in almost equal proportions. 
On the contrary, the number of atoms with bcc environment is negligible. 
Therefore, we have analized the results in terms of crystalline
(fcc+hcp) and icosahedral structures only.
In addition, as clusters of four-atom tetrahedra (i.e. regular
face-sharing tetrahedra) have been also proposed
as the basic building blocks in amorphous
metals, we also analize them.
We define a regular tetrahedron as
four neighbor atoms whose relative distances
differ by less than 0.55 \% \cite{mik}.

During the simulated quench, the heat capacity $C_P$ presents a
clear discontinuity at $T_g \sim 750$ K, which agrees well with the 
experimental glass transition temperature estimated for pure nickel \cite{kim}.
Although the drop in $C_P$ is rather
smooth because of the fast quenching rates, it is generally
accepted that this discontinuity is associated with the arrest
of structural relaxation, which prevents the system from
visiting all the possible configurational states and
is the landmark of the glass transition.
Thus, it can be said that pure nickel exhibits a glass transition.
As expected, $T_g$ increases moderately with the quenching rate.
It is important to stress that a slower quench at $7 \times 10^{12}$ K/s
resulted in a complete crystallization. Therefore, we are studying
the slowest rates which result in an amorphous state.
Fig. \ref{fig1} compares the radial and angular
distribution functions of the liquid with those of
one of the amorphous structures (excluding as central atoms
those with crystalline environment).
It is clear that they are almost identical.
The double second peak in $g(r)$, characteristic of amorphous metals,
is clearly visible but, contrarily to previous proposals \cite{tsu}, 
it is not related with icosahedra (which are very scarce, see below).
In fact, it is not even characteristic of the amorphous state,
since it also appears in liquid configurations
(relaxed to the closest local minimum).

Fig. \ref{fig2} shows the final number 
of atoms with crystalline and icosahedral environments
for all the quenches.
Two features are worth noting. 
First, the final number of crystalline atoms is proportional
to the total quench time.
This means that the rate of growth of crystalline order 
during the quench
is approximately independent of the quenching rate.
Second, the number of atoms with icosahedral environment
is small at all temperatures and it even decreases
during the slowest quenches.
The size of the tetrahedra clusters (not shown) also decreases.
This is in marked contrast with pair-potential simulations, in which
icosahedra and tetrahedra clusters grow at the quench and only
disappear during crystallization \cite{yon,jun2,mik}.
Typically, free (gas phase) icosahedral clusters are more stable 
than pieces of
fcc or hcp structures, due to the higher coordination on their
surface, so that a crystalline cluster needs to grow beyond some
critical size to become energetically favorable \cite{mar}.
This is still true with the embedded-atom potential for 
free nickel clusters, but not for clusters immersed in
a liquid or amorphous matrix, which have 
$\sim 30$ meV/atom more energy than crystalline clusters.
On the contrary, in pair potential models, atoms with icosahedral
environment have the lowest energy \cite{yon,tsu}, 
so that more abundant \cite{kon} and more stable \cite{yon} 
icosahedra are formed with slower quenches.

To address the question of how much aggregation
exists among the atoms with crystalline environment,
Fig. \ref{fig3} shows the three-dimensional structure
of some of the 6912-atom samples produced at different quenches.
Only atoms with crystalline environment are shown,
and clusters of them are clearly visible.
These clusters, if not perfectly crystalline, are regions
with good local order.
Larger crystalline clusters are observed for the slowest quenchings.
In agreement with our results,
it has been observed very recently that amorphous nickel consists
of regions with only short range order, together with regions of
nascent crystallinity \cite{bal}.

In order to study the relative stability of the glasses produced
at different quenching rates, we heated slowly three samples,
now with 856 atoms because the slow heating rates 
require more computational effort. 
These samples had been quenched at different
cooling rates and had correspondingly different proportions
of crystalline atoms.
The same heating rate of $3 \times 10^{12}$ K/s was used for all
three samples.
Fig. \ref{fig4} illustrates such reheating, showing a
sudden drop in energy when the samples crystallize.
It is clear that the more slowly cooled (and more crystalline) 
samples are less stable, crystallizing at a lower temperature.
We repeated the same procedure for three more samples, 
with the same result.
A visual inspection shows that crystallization occurs by
the growth of crystallites initially present.
Our results are again in contrast with those for pair potentials
\cite{yon,jun2}, but they are in agreement with the 
experimental observation that the most slowly cooled side
of a metallic glass ribbon crystallizes first \cite{kos}.
In fact, crystallization in metallic glasses
happens frequently by the growth of quenched-in
nuclei, which are more numerous after slower quenches \cite{kel}.

From the above results and from the slower simulated quenches, 
which resulted in complete crystallization,
the picture which emerges for the formation and the structure 
of pure amorphous metals is particularly simple.
As the liquid is cooled down, the rate of nucleation and growth of
crystallites reaches a maximum, determined by the increasing
difference in free energies and the decreasing rate of the dynamics.
This maximum approximately coincides with the glass
transition, because the arrest of the dynamics is also the
origin of the latter.
The slower the quenching rate, the larger the crystallites,
which grow immersed in an amorphous matrix
with essentially the same structure of the liquid.
The crystallites are well separated from each other
until the proportion of crystalline atoms is $\sim 30$\%.
If this limit is not reached, no clear signs of crystallization appear,
neither in the heat capacity nor in the resulting
pair correlation function.
Above this limit, the crystallization proceeds by the growth of
the preexisting crystallites, accelerating
because of their mutual interaction \cite{and,acc}, and producing
a clear energy discontinuity.
The resulting structure is then an aggregate of space-filling 
and well formed crystallites, 
with a very different pair correlation function.

In conclusion, we have found that
in pure amorphous metals there are no special structural motifs,
as in the icosahedral or polycluster models \cite{gas,bak}.
Instead, there is a continuous change
from liquid to crystal, arrested by the fast quench.
But, despite the absence of a special structure, this system
shows the most characteristic properties of the glass
transition, and of the amorphous solid state.
In this sense, pure metals are the
simplest glasses, showing only their essential kinetic and
thermodynamic properties (deriving from the arrest of the 
relaxation dynamics), but devoid of complex structural changes.
As such simplest systems, pure metals might help to clarify
many of the subtle properties of glasses.

We thank Mar\'{\i}a Aguilar for her help in implementing
the embedded atom method, and we acknowledge useful
discussions with 
M. A. Ramos, J. M. Rojo and M. A. Gonz\'{a}lez. 
This work was supported by Spanish DGICyT, under grant PB95-0202.


\begin{figure}[h]
\caption{
Radial (main figure) and angular (inset) distribution functions of the
liquid (dashed) and of one of the amorphous samples, 
quenched at $3\times 10^{13}$ K/s (solid).
In the latter, to show the structure of the 
pure amorphous matrix, only the distribution functions of 
the amorphous atoms (with non-crystalline environment, see text)
with all their neighbours are represented. 
In both cases, the structure was first 
relaxed to the closest energy minimum in order to avoid the 
vibrational noise. The slight shift between both curves in $g(r)$ is
mainly due to thermal expansion.}
\label{fig1}
\end{figure}

\begin{figure}[h]
\caption{
Number of atoms with crystalline (circles)
and icosahedral (diamonds) environment in the amorphous samples with 6912 atoms,
quenched from the liquid at different rates. $t$ is the
total time spent for the quench, from 2000 K to 100 K. The liquid configuration
is
represented at $t=0$. The lines are a linear fit and a guide to the eye.
Notice the different scales.}
\label{fig2}
\end{figure}

\begin{figure}[h]
\caption{
Simulation cells with the final configuration of samples 
quenched at different rates. Only atoms with crystalline environments
are shown. Quenching rates decrease from {\it a} to {\it d}.
{\it a}: $10^{14}$ K/s, {\it b}: $5 \times 10^{13}$ K/s, {\it c}: $3 \times
10^{13}$ K/s,
{\it d}: $10^{13}$ K/s.}
\label{fig3}
\end{figure}

\begin{figure}[h]
\caption{
Heating and crystallization of three 
amorphous samples, previously quenched at different rates. 
The numbers indicate the amount of crystalline atoms existing in 
the relaxed samples before heating.}
\label{fig4}
\end{figure}

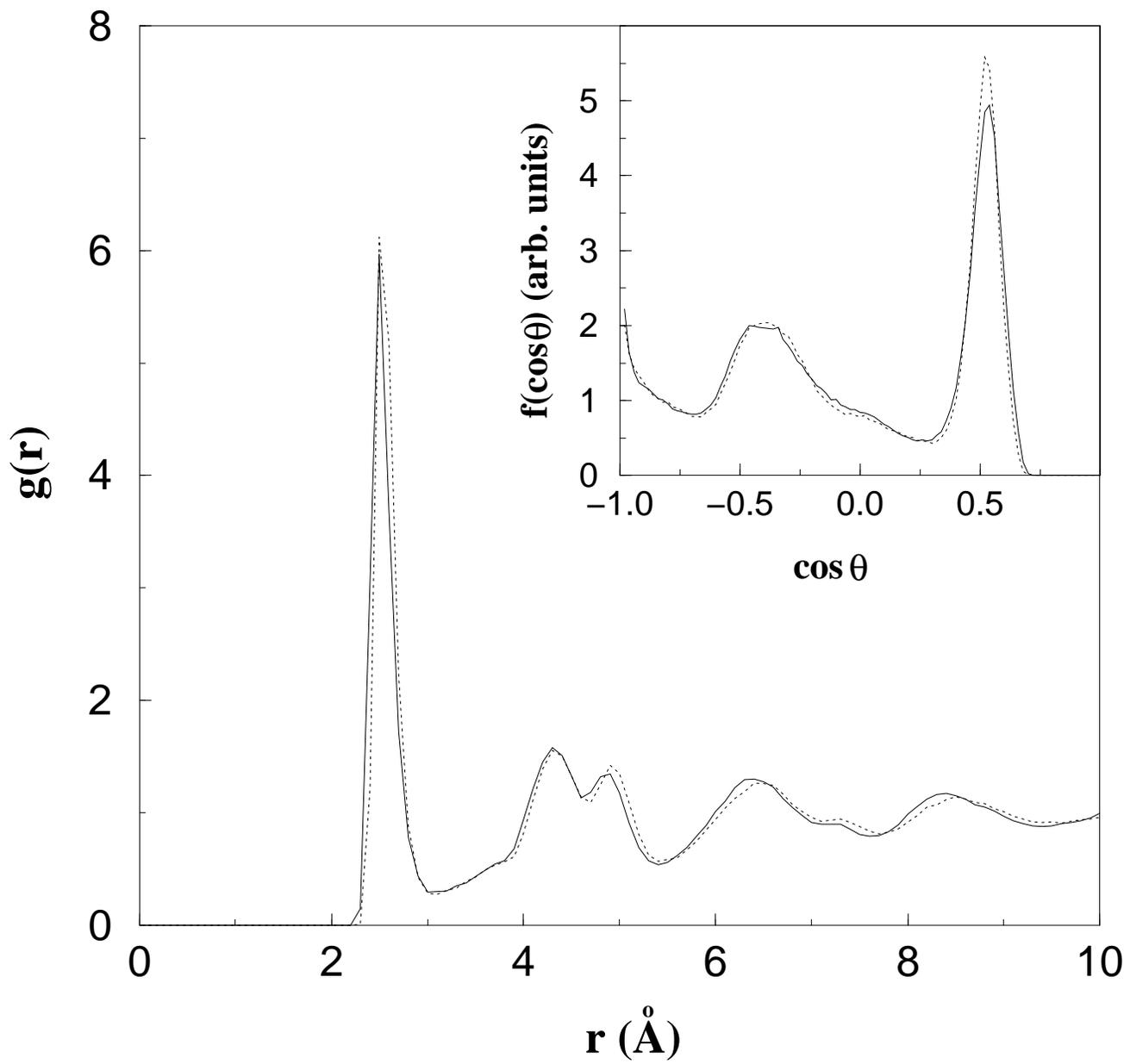
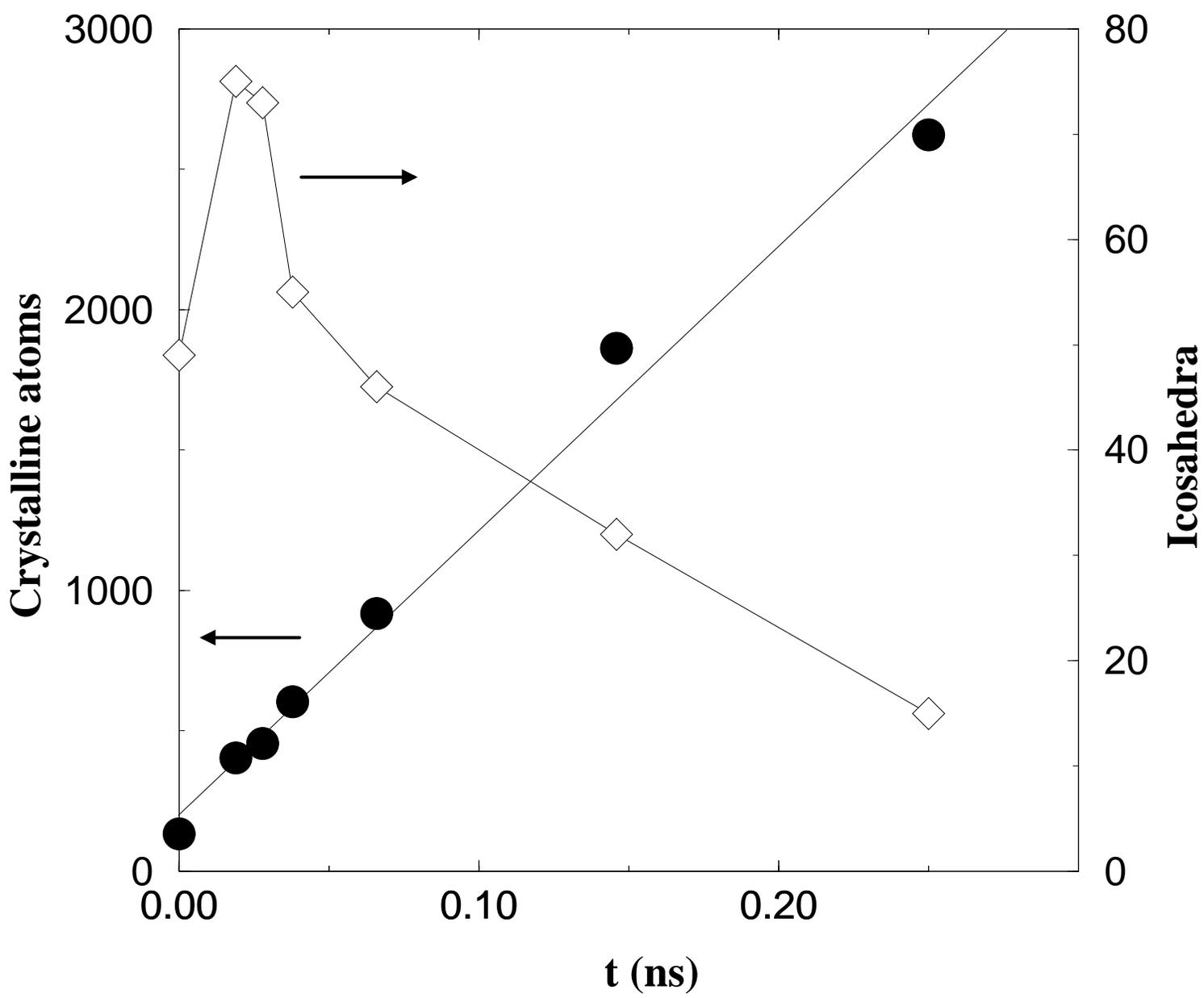
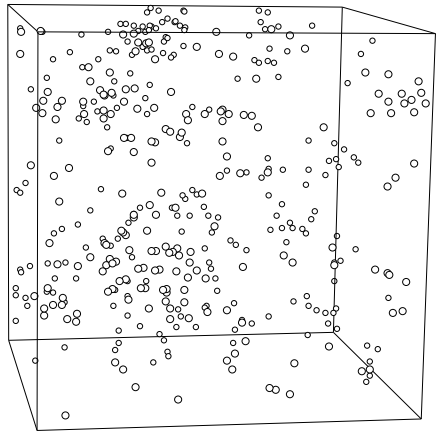
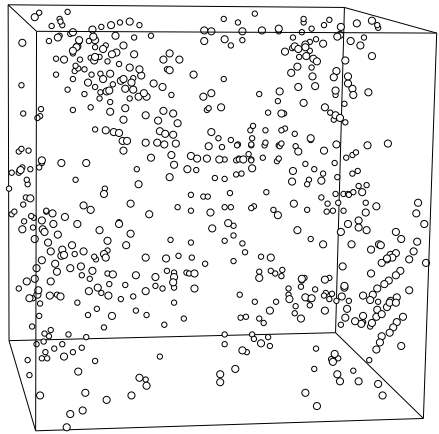
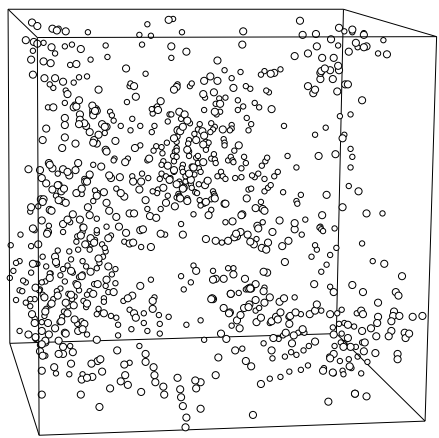
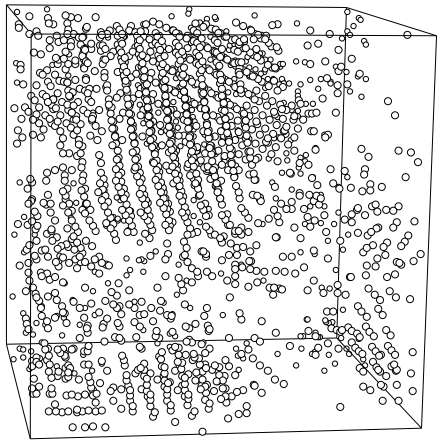
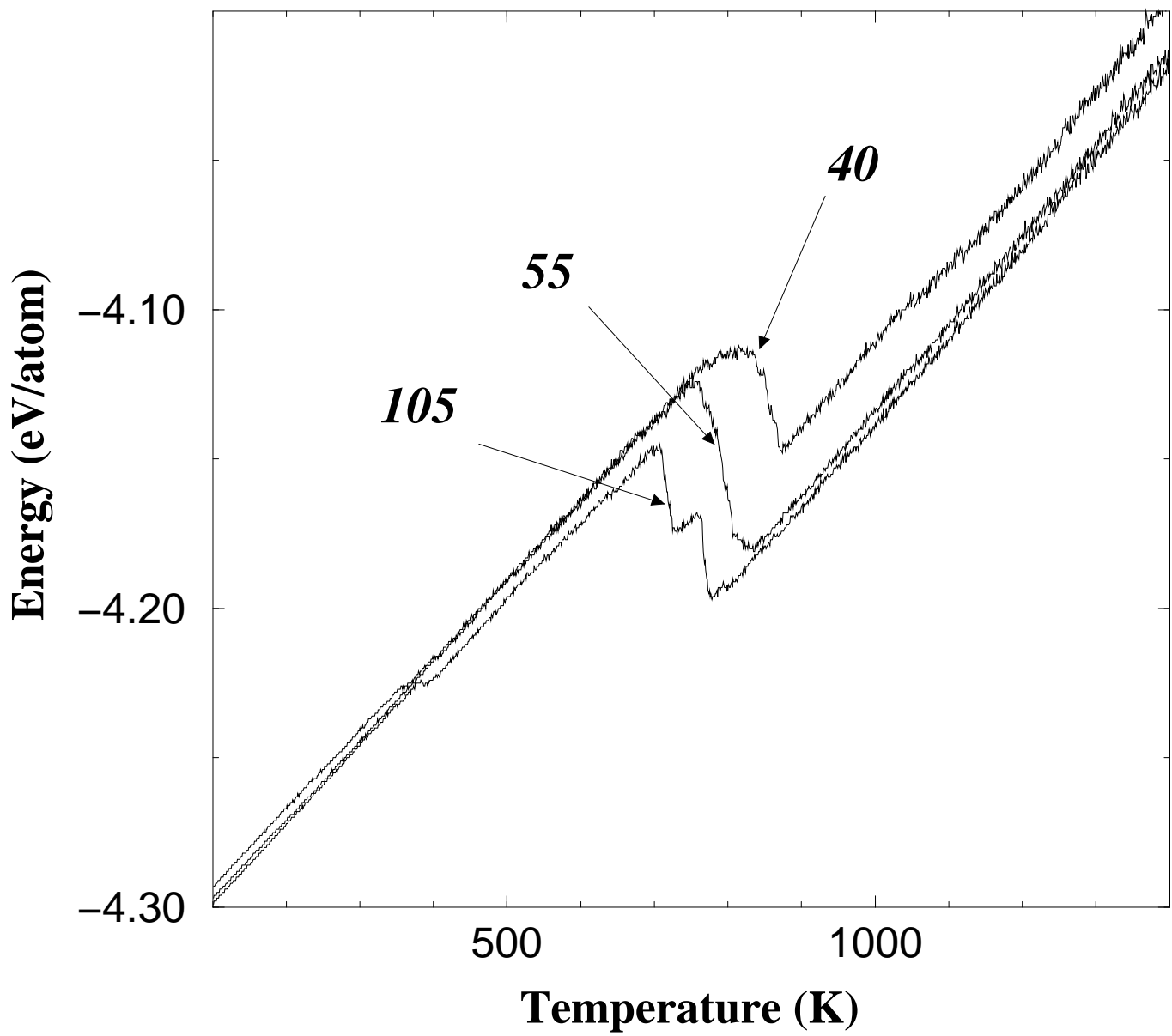
\end{document}